\documentclass [12pt] {iopart}

\begin{document}

\title
{Rotation of quantum liquid without singular vortex lines}

\author{Yakov GREENBERG$^1$ and Vladimir ZELEVINSKY$^2$}

\address {$^{1}$ Department of Physics and Techniques, Novosibirsk State Technical University, 20 K.
Marx Ave., Novosibirsk 630092, Russia}
\address {$^{2}$NSCL and Department of Physics and Astronomy,
Michigan State University, East Lansing, MI 48824-1321 USA}
\ead{greenbergy@online.nsk.su}

\date{\today}

\begin{abstract}

The operator equations for quantum hydrodynamics are discussed and solved
in a simple cylindrical geometry. We find a solution with the velocity curl
``frozen" into a density of the liquid in the absence of singular vortex lines.
The spectrum of small oscillations around this solution is found as a
generalization of the standard phonon spectrum, and the stability
of the solution is demonstrated.

\end{abstract}
\section{Introduction}

With the development of physics of cooled atoms in traps, the studies of
quantum rotation in a many-body system have revived and entered a new stage.
In superfluid helium-4, as it is well known \cite{andronik66,donnelly91},
the vortices with quantized circulation emerge at a certain critical
cranking velocity. At sufficiently fast rotation, the vortex lines form
a lattice that on average imitates the rigid-body rotation \cite{feynman55}.
Analogous vortices appear in penetration of the magnetic field into
superconductors of second kind \cite{abrikosov57}. Roughly speaking,
the first vortex can be born by a perturbation violating the angular momentum
conservation at the angular velocity corresponding to the crossing
of single-particle levels corresponding to the zero and non-zero
angular momentum per particle. A nuclear analog is the so-called back-bending
that occurs also at a band crossing in the rotating frame \cite{bengtsson79}.
Similar phenomena are known for molecules \cite{tyng07}, predicted for
neutron stars \cite{glendenning97}, and important for some quantum
field-theoretical models \cite{obukhov97}.

Our study is also partly motivated by the recent explosion of
activity in experimental and theoretical studies of superfluidity
and other quantum phenomena in trapped ultracold atomic systems
\cite{fetter09,onofrio00,marago00} and quantum Hall superfluids
\cite{MacDonald01,Abolfath03}. Also, there have been many
experimental studies of the flow properties of superfluids
confined in porous media such as Vycor glass and containers packed
with fine powder \cite{kim04,cao86,yamamoto04,bossy08}.

The question of a possibility for a quantum liquid to be in a
state of rotational excitation without forming individual vortices
or a vortex lattice is still debatable. Such rotation should have
a moment of inertia close to that of a rigid body. A small quantum
system, such as a medium or heavy atomic nucleus, clearly displays
numerous rotational bands built on various intrinsic
configurations. A Fermi system without pairing correlations should
have, with semiclassical accuracy, the moment of inertia
corresponding to the rigid-body rotation for a given density
distribution \cite{rockmore60}. In agreement with the substantial
presence of pairing, the nuclear moment of inertia is considerably
smaller than that of a rigid body but still much greater than that
of a perfect liquid \cite{belyaev59,migdal59}. Rotation of an
atomic system with and without a quantized vortex was considered
in \cite{corro07} (see also references therein). The evidence for
the so-called Landau state with identically vanishing curl of the
superfluid velocity was presented in \cite{kojima71}. Especially
rich spectrum of rotational phases is predicted and partially
observed in supefluid helium-3 \cite{hakonen87,lounasmaa99}, where
the orbital momentum is strongly coupled with spin structures. New
types of rotation are expected at high pressure
\cite{babaev04,babaev05} and for crystallizing quantum liquids
\cite{reimann06}. It is also worthwhile to mention here the
fractional quantum Hall effect where the vorticity of Laughlin's
state is proportional to the local density of electronic fluid
\cite{Stone90,Setlur04}.

The statement that a quantum liquid rotates necessarily with
emergence of quantized vortex lines seems to be too restricting.

In the present article we consider the quantum hydrodynamics in the original
form developed by Landau \cite{landau41}. With the exact solution of quantum
operator equations of motion we come to the type of rotation that is not related
to the formation of singular vortex lines (the vortex solution is possible
as well). We show also that the new solution is stable with respect to small
oscillations and find explicitly the spectrum of such oscillatory modes.

\section{Quantum hydrodynamics in operator formulation}

The equations of quantum hydrodynamics for a system of identical
Bose-particles of mass $m$ were formulated by Landau in
terms of the field operators of density, $\rho({\bf x})=mn({\bf x})$, and
current density ${\bf j}({\bf x})$,
\begin{equation}
n({\bf x})=\sum_{a}\delta({\bf r}_{a}-{\bf x}),             \label{1}
\end{equation}
\begin{equation}
{\bf j}({\bf x})=\frac{1}{2m}\,\sum_{a}[{\bf p}_{a}, \delta({\bf r}_{a}-{\bf
x})]_{+},                                             \label{2}
\end{equation}
where the subscript $a$ labels single-particle variables, and $[...\,,\,...]_{+}$
means the anticommutator. The quantization is performed through the commutation relations,
\begin{equation}
[n({\bf x}),n({\bf x}')]=0,                          \label{3}
\end{equation}
\begin{equation}
[{\bf j}({\bf x}),n({\bf x}')]=-i\,\frac{\hbar}{m}\,n({\bf x}) \nabla_{{\bf
x}}\delta({\bf x}-{\bf x}'),                       \label{4}
\end{equation}
and
\begin{equation}
    \left[ {j_i (\textbf{x}),j_j (\textbf{x}')} \right] =
    i\,\frac{\hbar}{m}\, \varepsilon _{ijk} \left( {{\mathop{\rm curl}\nolimits}\, \textbf{j}(\textbf{x})}
     \right)_k \delta (\textbf{x} -\textbf{x}').   \label{5}
\end{equation}

The microscopic velocity field ${\bf v}({\bf x})$ can be
defined (at the points where the density $n$ differs from zero) as
\begin{equation}
{\bf v}({\bf x})=\frac{1}{2}\,\Bigl[\frac{1}{n({\bf x})}, \,{\bf j}({\bf
x})\Bigr]_{+}.                                    \label{6}
\end{equation}
It satisfies the commutation relations
\begin{equation}
[{\bf v}({\bf x}),n({\bf x}')]=-i\,\frac{\hbar}{m}\,\nabla_{{\bf
x}}\,\delta({\bf x}-{\bf x}'),                     \label{7}
\end{equation}
and
\begin{equation}
[v_{i}({\bf x}),v_{j}({\bf x}')]=i\frac{\hbar}{m}\,\epsilon_{ijk}\,
\frac{1}{n({\bf x})}\,({\rm curl}\,{\bf v}({\bf x}))_{k}\,
\delta({\bf x}-{\bf x}').                        \label{8}
\end{equation}
As a consequence of eq. (\ref{7}), the density operator commutes with the curl
of velocity,
\begin{equation}
[n({\bf x}),{\rm curl}\,{\bf v}({\bf x}')]=0.   \label{9}
\end{equation}
The whole algebra can be simply postulated or derived in terms of
the underlying particle variables, either in configuration space
\cite{landau41} (the right hand side of eq. (\ref{8}) is given there with
an opposite sign), or in the secondary quantized form \cite{geilikman54}.

There are statements in the literature, for example
\cite{frohlich67,garrison70}, that quantum hydrodynamics formulated in such a
way contains intrinsic contradictions. Indeed, with formal manipulations by
commutation relations, one can construct the states, where the expectation
value of $n({\bf x})$ is negative and very large by absolute magnitude.
However, this is not a physical difficulty. Thinking of a real quantum liquid,
we should have in mind the Hamiltonian of the type
\begin{equation}
H=\frac{m}{2}\,\int d^{3}x\,{\bf v}({\bf x}){n}({\bf x}) {\bf v}({\bf x})
+W\{n\},                                          \label{10}
\end{equation}
where the functional $W$ depends on density and its spatial gradients.
Since the liquid consists of quantum particles, this functional contains
the well known quantum potential $W_{q}$,
\begin{equation}
W=W_{q}+\overline{W}, \quad W_{q}=\,\frac{\hbar^{2}}{8m}\,\int d^{3}x\,
\frac{(\nabla n)^{2}}{n}.                      \label{11}
\end{equation}
Similarly to the centrifugal barrier in the radial Schr\"{o}dinger equation,
the term $W_{q}$ does not allow the system to penetrate into the regions of
$n<0$. Such ``fall onto the center" is possible only if the interaction
$\overline{W}$ has a stronger singularity at $n\rightarrow 0$. We can
conclude that the growth of quantum fluctuations at $n\rightarrow 0$ prevents
the transition into unphysical states.

Being interested in the spectrum of low-lying elementary excitations in
superfluid $^{4}$He (phonons and rotons), Landau considered in Ref.
\cite{landau41} only irrotational motion, still mentioning the possibility of
vortex motion in some quantum states which are to be separated by an energy gap
from the ground state. Geilikman \cite{geilikman54} linearized the equations of
motion for weak excitations above the uniform background (${\bf v}^{\circ}=0,\;
n^{\circ}=\,{\rm const}$), found the well known Bogoliubov phonon spectrum
\cite{bogoliubov47}, and showed that the modes with ${\rm curl}\,{\bf v}\neq 0$
appear only in higher orders as a result of phonon anharmonicity. Below we
construct a class of solutions, where both, classical ($c$-number) and quantum
(operator), parts of ${\rm curl}\,{\bf v}$ are present.

\section{``Frozen" rotation}

Let us define an axial vector operator $\vec{\Gamma}({\bf x})$
that, according to eq. (\ref{9}), commutes with $n({\bf x})$,
\begin{equation}
{\rm curl}\,{\bf v}({\bf x})=n({\bf x})\vec{\Gamma}({\bf x}). \label{12}
\end{equation}
Using our operator algebra, we can derive the commutation
relations of the components $\Gamma_{i}$ among themselves and with
components of velocity, $v_{j}$. The simplest non-trivial
solutions compatible with the operator algebra correspond to the
states of the ``frozen" rotation \cite{GZ79}, when $\vec{\Gamma}$ is a
constant $c$-number vector, let say along the $z$-axis, while the
density and velocity fields do not depend on $z$. In this
case
\begin{equation}
    [{\rm curl}\,{\bf{v(x)}},H]=-\frac{i\hbar}{2}\,\vec{\Gamma}\,
    {\rm div}[{\bf v(x)}n({\bf x})+n({\bf x}){\bf v(x)}]=
    i\hbar\vec{\Gamma}\,\frac{\partial n({\bf x})}{\partial t}. \label{13}
\end{equation}
Therefore, for time-independent density, the quantity
$\texttt{curl}\,\textbf{v}(\textbf{x})$ is the constant of
motion that characterizes the stationary states of the system.

We consider a planar motion ${\bf v}({\bf r})$ in a circular
cylinder of radius $R$ with longitudinal axis $z$, $v_{z}=0$. The
transverse components of the velocity field satisfy, according to
eqs. (\ref{8}) and (\ref{12}), the commutation relations similar to
those in a magnetic field,
\begin{equation}
[v_{x}({\bf r}),v_{y}({\bf r}')]=\,\frac{i\hbar}{m}\,\Gamma\delta({\bf r}-{\bf
r}'),                                                   \label{14}
\end{equation}
where ${\bf r}=(r,\varphi)$ is the coordinate vector in the $(xy)$-plane. The
operator equations of motion for the velocity field, on this class of states,
are given by
\begin{equation}
[v_{i}({\bf r}),H]=-\,\frac{i\hbar}{m}\,\frac{\partial}{\partial x_{i}}\left\{
\frac{mv^{2}({\bf r})}{2}\,+\,\frac{\delta W}{\delta n({\bf r})}\,\right\}+
i\hbar\Gamma\epsilon_{ijz}\,\frac{1}{2}\,[n({\bf r}),v_{j}({\bf r})]_{+}.
                                                          \label{15}
\end{equation}
Here the density $n({\bf r})$ has to be expressed in terms of
${\rm curl}\,{\bf v}$ due to eq. (\ref{12}).

We will look for the solution of the set (\ref{14},\ref{15}) as
small oscillations above the background characterized by the
classical fields of velocity, ${\bf v}^{\circ}({\bf r})$, and
density, determined by ${\rm curl}_{z}\, {\bf v}^{\circ}({\bf
r})=\Gamma n^{\circ}({\bf r})$. We limit ourselves by the axially
symmetric case,
\begin{equation}
v^{\circ}_{\varphi}({\bf r})=u(r), \quad v^{\circ}_{r}=0, \quad n^{\circ}({\bf
r})=\nu(r).                                         \label{16}
\end{equation}
Of course, the angular momentum of the system will be directed along the axis
but, due to the azimuthal symmetry, there is no need to make a transformation
to a rotating frame and introduce the Lagrangian multiplier of the angular
velocity. The similar situation is known also in nuclear physics where the
rotational motion around the symmetry axis does not require the cranking
procedure. The difference here is that we find collective excited states
instead of the more or less random sequence of single-particle excitations
carrying the angular momentum projection onto the symmetry axis in a nucleus
rotating around the same axis.

The background state, as follows from eq. (\ref{15}), should
satisfy the equilibrium condition,
\begin{equation}
\frac{mu^{2}}{r}\,=\,\frac{d}{dr}\left(\frac{\delta W}{\delta
n}\right)_{n=\nu(r)}.                             \label{17}
\end{equation}
In our geometry,
\begin{equation}
\frac{du}{dr}\,=-\,\frac{u}{r}\,+\Gamma \nu(r).       \label{18}
\end{equation}
This leads to the general form of the background velocity field,
\begin{equation}
u(r)=\,\frac{1}{r}\left[K+\Gamma\int_{0}^{r}dr'\,r'\nu(r')\right]=
=\,\frac{1}{r}\left[K+\,\frac{\Gamma}{2\pi}\,N(r)\right], \label{19}
\end{equation}
where $K$ is a constant and $N(r)$ is a number of particles inside the radius
$r$, $0<r<R$; here and later all macroscopic quantities are taken per unit
length of a container (see also Appendix).

Our general solution (\ref{19}) represents a superposition of a vortex line of
intensity $K$ and rigid-body rotation of cylindrical layers with a local
angular velocity $\Omega(r)=\Gamma\nu(r)/2$. The angular momentum of the liquid
$L=L_{z}$ can be found as
\begin{equation}
L=m\int d^{2}r\,\nu(r)ru(r).                          \label{20}
\end{equation}

Using eq. (\ref{19}), we derive
\begin{equation}
L=mNK+\,\frac{m\Gamma}{4\pi}\,N^{2},                  \label{21}
\end{equation}
where $N=N(R)$ is the total particle number. In the absence of $\Gamma$, as usual,
the bosonic quantization $L=N\hbar s$, where $s$ is an integer, determines
$K_{s}=s\hbar/m$. The additional part of the angular momentum does not carry
any singularity at $r\rightarrow 0$.

For a vortex-free case, $K=0$, we have from  $L=N\hbar s$
\begin{equation}
    \Gamma=\frac{4\pi\hbar}{mN}s,                      \label{22}
\end{equation}
which corresponds to rotation with local linear velocity
$v_b(r)=\Gamma\nu(r)r/2.$

\section{Solutions with and without a vortex}

To move further, we assume for simplicity that the functional $\overline{W}$
does not contain gradients of $n$ and can be expanded around the unperturbed
density of non-rotating liquid, $n_{0}$,
\begin{equation}
\overline{W}=\overline{W}\{n_{0}\}+\,\frac{1}{2}\,m^{2}g\int
d^{2}r[n(r)-n_{0}]^{2}.                               \label{23}
\end{equation}
With this functional, the balance equation (\ref{17}) takes the
form (see Appendix for derivation)
\begin{equation}
\frac{mu^{2}}{r}\,=m^{2}g\,\frac{d\nu}{dr}\,+\,\frac{\hbar^{2}}{8m}\,
\frac{d}{dr}\,\left[\frac{1}{\nu^{2}}\,\left(\frac{d\nu}{dr}\right)^{2}\,-
\,\frac{2}{\nu r}\,\,\frac{d}{dr}\left(r\,\frac{d\nu}{dr}\right)\right].
                                                   \label{24}
\end{equation}

The compressibility $g$ defines the speed of sound,
$c=\sqrt{mgn_{0}}$. It is convenient to introduce dimensionless
variables and express velocities in units of $c$, energies in
units $mc^{2}$, lengths in units of $r_{0}=\hbar/mc$, and density
in units of $n_{0}$. Then the only remaining parameter is $\Gamma$
[in usual units $\Gamma\Rightarrow \hbar n_{0}\Gamma/(mc^{2})$].
For typical quantum liquids, the length $r_{0}$ corresponds to
atomic distances, while the macroscopic container has $R\gg 1$. At
the same time, for reasonable angular velocity, the rotation is
subsonic, $\Gamma R<1$, which means $\Gamma\ll 1$ in reduced
units.

As follows from eq. (\ref{19}), we can have two types of solutions for
eq. (\ref{24}) with a regular behavior of density $\nu(r)$ at the axis,
$r\rightarrow 0$. For the vortex line solution, $K\neq 0$, at small distances
from the axis the velocity is $u\approx K/r$, while the density goes to zero,
$\nu(r)\propto r^{2|K|}+ {\cal O}(r^{4|K|+2})$, and its behavior is determined
by the quantum term (\ref{11}). Here $\Gamma$ influences only higher terms of
the expansion of $\nu(r)$, the radius of the vortex core does not noticeably
depend on $\Gamma$ and has the magnitude of the order of one (the distance
where we can equate contributions to the energy from compressibility and from
quantum effects). Such results for quantized vortices are well known
\cite{andronik66,ginzburg58}; they are modified by rotation only at large
distances from the center. Below we will be interested in the solutions of the second
type, where the vortex singularity is absent, $K=0$.

At $K=0$, the velocity and density satisfying eqs. (\ref{24}) and (\ref{19})
can be expressed by a well converging series in powers of a parameter
$\xi=\nu_{0}\Gamma^{2}R^{2}$:
\begin{equation}
\nu(r)=\nu_{0}\left\{1+\,\frac{\xi}{8}\,\frac{r^{2}}{R^{2}}\,+\,
\frac{\xi^{2}}{128}\,\frac{r^{4}}{R^{4}}\,+\,...\right\},   \label{25}
\end{equation}
\begin{equation}
u(r)=\,\frac{1}{2}\,\nu_{0}\Gamma r\left\{1+\,\frac{\xi}{16}\,
\frac{r^{2}}{R^{2}}\,+\,\frac{\xi^{2}}{384}\,\frac{r^{4}}{R^{4}}+
\,...\right\}.                                         \label{26}
\end{equation}
The central density $\nu(0)\equiv \nu_{0}$ is determined by the particle
number conservation being related to the unperturbed value
$n_{0}=2u(R)/(R\Gamma)\equiv 1$ through a similar series,
\begin{equation}
\nu_{0}^{-1}=1+\,\frac{\xi}{16}\,+\,\frac{\xi^{2}}{384}\,+\,... \label{27}
\end{equation}

It is easy to estimate that in such a state the contributions to energy per
particle from the kinetic term and from compressibility are, by order of
magnitude, $\sim \Gamma^{2}R^{2}$ and $\sim\Gamma^{4}R^{4}$, respectively,
whereas the quantum potential brings in only a small contribution
$\sim\Gamma^{2}$. The angular momentum, eq. (\ref{21}), equals $L=N\hbar s$,
where $s=\Gamma R^{2}/4$; at $\Gamma R\sim 1$, this value $s\gg 1$. In the case
of the external cranking with angular velocity $\Omega$, the state with a
certain value $\Gamma\neq 0$ is getting energetically favorable at a critical
value
\begin{equation}
\Omega_{c}(\Gamma)=\,\frac{E(\Gamma)-E(0)}{L(\Gamma)}\,\approx\,\frac{\Gamma}{4}.
                                                         \label{28}
\end{equation}
For the minimum value $\Gamma_{{\rm min}}=\Gamma(s=1)\approx 4/R^{2}$, we have
$\Omega_{c}^{{\rm min}}\approx R^{-2}$. This type of rotation can emerge at a
smaller (by a factor of $\ln R$) angular velocity than the creation of usual
quantized vortices. However, the probability of such an excitation, for example
by an accidental nonuniformity of the walls of the container \cite{putterman72},
is considerably lower, while the observation at $\Gamma R<1$ is difficult.

\section{Small amplitude oscillations}

Now we will construct the low-lying excited states corresponding to small
oscillations of the density and transverse velocity around the classical
solution [eqs. (\ref{25}-{\ref 27})]. We introduce the exciton operators,
$\hat{{\bf v}}({\bf r})\equiv{\bf v}-{\bf u}$ and $\hat{n}({\bf r})\equiv
n-\nu$. The linearization of the equations of motion (\ref{15}) in the model
(\ref{23}) gives
\begin{equation}
[\hat{{\bf v}},H]=-i\vec{\nabla}({\bf u}\cdot\hat{{\bf v}}+\hat{P})+i[({\bf u}
\hat{n}+\hat{{\bf v}}\nu)\times\vec{\Gamma}],               \label{29}
\end{equation}
or, in cylindrical coordinates,
\begin{equation}
[\hat{v}_{r},H]=-i\,\frac{\partial}{\partial r}(u\hat{v}_{\varphi}+\hat{P})
+i\Gamma(\hat{v}_{\varphi}\nu+u\hat{n}),                    \label{30}
\end{equation}
and
\begin{equation}
[\hat{v}_{\varphi},H]=-\,\frac{i}{r}\,\frac{\partial}{\partial\varphi}\,
(u\hat{v}_{\varphi}+\hat{P})-i\Gamma\hat{v}_{r}\nu.       \label{31}
\end{equation}
Here the operator was introduced,
\begin{equation}
\hat{P}=\hat{n}-\,\frac{1}{4}\,\left\{\nabla^{2}\left(\frac{\hat{n}}{\nu}\right)
-\,\frac{\vec{\nabla}\nu}{\nu}\cdot\vec{\nabla}\left(\frac{\hat{n}}{\nu}\right)
\right\},                                                  \label{32}
\end{equation}
and, according to eq. (\ref{12}),
\begin{equation}
\Gamma\hat{n}=\,\frac{1}{r}\left\{\frac{\partial}{\partial r}\,
(r\hat{v}_{\varphi})-\,\frac{\partial\hat{v}_{r}}{\partial\varphi}\right\}.
                                                         \label{33}
\end{equation}
The continuity equation,
\begin{equation}
[\hat{n},H]=-i\,\frac{u}{r}\,\frac{\partial\hat{n}}{\partial\varphi}\,-\,
\frac{1}{r}\left\{\,\frac{\partial}{\partial r}(r\nu\hat{v}_{r})+i\nu
\,\frac{\partial\hat{v}_{\varphi}}{\partial\varphi}\right\},  \label{34}
\end{equation}
follows from the equations of motion. Finally, with use of eq. (\ref{18}), we
can write down the equations of motion in a symmetric form,
\begin{equation}
[\hat{v}_{r},H]+i\,\frac{u}{r}\,\frac{\partial\hat{v}_{r}}{\partial\varphi}\,
-2i\,\frac{u}{r}\,\hat{v}_{\varphi}=-i\,\frac{\partial\hat{P}}{\partial r},
                                                            \label{35}
\end{equation}
\begin{equation}
[\hat{v}_{\varphi},H]+i\,\frac{u}{r}\,\frac{\partial\hat{v}_{\varphi}}{\partial\varphi}\,
+i\Gamma\nu\,\hat{v}_{r}=-\,\frac{i}{r}\frac{\partial\hat{P}}{\partial\varphi}.
                                                            \label{36}
\end{equation}

At azimuthal symmetry of the ground state, the elementary Bose-excitations
can be characterized by a certain angular momentum $L_{z}=\hbar \ell$. The quantization
goes with expansion of our fields over corresponding creation, $A^{\dagger}_{kl}$,
and annihilation, $A_{kl}$, operators, where the additional subscript $k$ labels
consecutive modes for given $\ell$. The expansion can be written as
\begin{equation}
\left(\begin{array}{c}
\hat{v}_{r}({\bf r}\\
\hat{v}_{\varphi}({\bf r})\\
\hat{n}({\bf r})\end{array}\right)=\sum_{\ell}\,\frac{e^{i\ell\varphi}}{\sqrt{2\pi}}
\left(\begin{array}{c}
\hat{v}^{(\ell)}_{r}(r) \\
\hat{v}_{\varphi}^{(\ell)}(r)\\
\hat{n}^{(\ell)}(r)\end{array}\right)=\sum_{k\ell}\,
\frac{e^{i\ell\varphi}}{\sqrt{4\pi}}
\left(\begin{array}{c}
\left\{R_{k\ell}(r)A_{k\ell}+R^{\ast}_{k-\ell}(r)A^{\dagger}_{k-\ell}\right\}\\
i\left\{\Phi_{k\ell}(r)A_{k\ell}-\Phi^{\ast}_{k-\ell}(r)
A^{\dagger}_{k-\ell}\right\}\\
i\left\{F_{k\ell}(r)A_{k\ell}-F^{\ast}_{k-\ell}(r)A^{\dagger}_{k-\ell}\right\}
\end{array}\right).                                  \label{37}
\end{equation}
Here the unknown radial form-factors of excitations,
$R_{k\ell},\,\Phi_{k\ell}$, and $F_{k\ell}$, are the matrix elements between
the ground state and excited one-quantum state with quantum numbers $k,\ell$
and energy $E_{k\ell}$. They satisfy the set of linear differential equations,
\begin{equation}
\left(E_{k\ell}-\ell\,\frac{u}{r}\right)R_{k\ell}+2\,\frac{u}{r}\,\Phi_{k\ell}
=\,\frac{dP_{k\ell}}{dr},                                      \label{38}
\end{equation}
\begin{equation}
\left(E_{k\ell}-\ell\,\frac{u}{r}\right)\Phi_{k\ell}+\Gamma\nu R_{k\ell}=\,\frac{\ell}{r}\,P_{k\ell},
                                                          \label{39}
\end{equation}
\begin{equation}
\Gamma F_{k\ell}=\,\frac{d\Phi_{k\ell}}{dr}\,+\,\frac{\Phi_{k\ell}}{r}\,-
\,\frac{\ell}{r}\,P_{k\ell}.                              \label{40}
\end{equation}
The right hand side quantities $P_{k\ell}$ are related to the matrix elements
of density:
\begin{equation}
P_{k\ell}=F_{k\ell}-\,\frac{1}{4}\,\left[\nabla_{\ell}^{2}-
\,\frac{1}{\nu}\,\frac{d\nu}{dr}\,\frac{d}{dr}\right]\,\frac{F_{k\ell}}{\nu},
                                                          \label{41}
\end{equation}
where the differential operator $\nabla_{\ell}^{2}$ is defined as
\begin{equation}
\nabla_{\ell}^{2}\equiv \,\frac{d^{2}}{dr^{2}}\,+\,\frac{1}{r}\,\frac{d}{dr}\,-
\,\frac{\ell^{2}}{r^{2}}.                           \label{42}
\end{equation}

To analyze this set of equations, it is convenient to introduce the notations
\begin{equation}
{\cal E}_{k\ell}(r)=E_{k\ell}-\ell\,\frac{u}{r}, \quad D_{k\ell}(r)=
{\cal E}_{k\ell}^{2}- 2\Gamma\,\frac{\nu u}{r},           \label{43}
\end{equation}
where ${\cal E}_{k\ell}(r)$ can be interpreted as ``excitation energy" in the
frame rotating together with the layer $r=$ const. The matrix elements of
the velocity components can be presented as
\begin{equation}
R_{k\ell}(r)=D_{k\ell}^{-1}\left({\cal E}_{k\ell}P'_{k\ell}-2\ell\,
\frac{u}{r^{2}}\,P_{k\ell}\right),                        \label{44}
\end{equation}
and
\begin{equation}
\Phi_{k\ell}(r)=D_{k\ell}^{-1}\left({\cal E}_{k\ell}\,\frac{\ell}{r}\,P_{k\ell}-
\Gamma \nu P'_{k\ell}\right);                           \label{45}
\end{equation}
here and below the prime indicates the radial derivative $d/dr$.
Returning to the definition (\ref{40}), we obtain a differential equation
for $F_{k\ell}$:
\[D_{k\ell}F_{k\ell}+\nu\nabla_{\ell}^{2}P_{k\ell}\]
\begin{equation}
=\left\{-\,\frac{\ell^{2}}{r^{2}}\left(\nu-
\,\frac{2u}{r\Gamma}\right)+\,\frac{\ell}{r}\left({\cal
E}'_{k\ell}\Gamma^{-1}-\,\frac{D'_{k\ell}} {D_{k\ell}}\,{\cal
E}_{k\ell}\right)\right\}P_{k\ell}-\left(\nu'-\nu\,\frac{D'_{k\ell}}
{D_{k\ell}}\,\right)P'_{k\ell}.                      \label{46}
\end{equation}
As seen from eq. (\ref{41}), the right hand side of (\ref{46}) is the third
order differential operator acting on $F_{k\ell}$; the coefficients of
the operator contain only small terms of the expansions (\ref{25}) and (\ref{26}).
Therefore we can act iteratively.

Here we will limit ourselves by the main order assuming
\begin{equation}
\nu\approx \nu_{0}\approx 1, \quad \frac{u}{r}\approx\,\frac{\Gamma}{2}, \quad
{\cal E}_{k\ell}\approx E_{k\ell}-\,\frac{1}{2}\,\ell\Gamma=\,{\rm const}.
                                                           \label{47}
\end{equation}
Then eq. (\ref{46}) is reduced to
\begin{equation}
\Bigl({\cal E}^{2}_{k\ell}-\Gamma^{2}+\nabla_{\ell}^{2}-\,\frac{1}{4}\,
\nabla_{\ell}^{4}\Bigr)F_{k\ell}^{(0)}(r)=0.                \label{48}
\end{equation}
The regular at the origin solution is a superposition of cylindrical functions
\begin{equation}
F_{k\ell}^{(0)}(r)=b^{(+)}_{k\ell}J_{\ell}(q^{(+)}_{k\ell}r)+
b^{(-)}_{k\ell}I_{\ell}(q^{(-)}_{k\ell}r),                     \label{49}
\end{equation}
where the wave numbers $q^{(\pm)}_{k\ell}$ are related to energy $E_{k\ell}$ by
a dispersion relation
\begin{equation}
{\cal E}_{k\ell}^{2}=\Gamma^{2}\pm q^{(\pm)2}_{k\ell}+\,\frac{1}{4}
\,q^{(\pm)4}_{k\ell},                                         \label{50}
\end{equation}
or, selecting the appropriate roots,
\begin{equation}
q^{(\pm)2}_{k\ell}=2\Bigl(\sqrt{1+{\cal E}^{2}_{k\ell}-\Gamma^{2}}\,\mp 1\Bigr).
                                                          \label{51}
\end{equation}
In the same approximation we obtain for the function (\ref{41}):
\begin{equation}
P^{(0)}_{k\ell}(r)=\,\frac{1}{4}\,\left\{b^{(+)}_{k\ell}q^{(-)2}_{k\ell}
J_{\ell}(q^{(+)}_{k\ell}r)-b^{(-)}_{k\ell}q^{(+)2}_{k\ell}
I_{\ell}(q^{(-)}_{k\ell}r)\right\}.                    \label{52}
\end{equation}
We came to the direct generalization of the standard phonon spectrum for the
case of the frozen vortex-like motion. There is now a small gap $\Gamma$ in the
phonon spectrum of transverse oscillations. The gap is determined by the
magnitude of curl ${\bf v}$.

The coefficients $b^{(\pm)}_{k\ell}$ of the superposition (\ref{49}),
as well as quantization of the eigenvalues ${\cal E}_{k\ell}$, are
determined by the normalization of the solutions and the boundary
conditions at $r=R$. For our problem of vibrations on the background of
a classical solution with almost constant density, the natural boundary
conditions in the accepted approximation are those of vanishing normal
components of the velocity, $R_{k\ell}$, and of the gradient of density,
$F'_{k\ell}$. A detailed discussion of boundary conditions can be found
in Ref. \cite{ziff77}. With the aid of eqs. (\ref{41},\ref{49}) and
(\ref{52}), we find, up to a normalization constant $C_{k\ell}$,
\begin{equation}
b^{(+)}_{k\ell}=C_{k\ell}I'_{\ell}(q^{(-)}_{k\ell}R), \quad b^{(-)}_{k\ell}=
-C_{k\ell}J'_{\ell}(q^{(+)}_{k\ell}R).                     \label{53}
\end{equation}
The energy quantization is given by $(z_{\pm}=q^{(\pm)}_{k\ell}R)$
\begin{equation}
J'_{\ell}(z_{+})I'_{\ell}(z_{-})\,\frac{z_{+}^{2}+z_{-}^{2}}{R^{2}}-\Gamma\,
\frac{\ell}{R^{3}}\Bigl[z_{-}^{2}J_{\ell}(z_{+})I'_{\ell}(z_{-})+
z_{+}^{2}I_{\ell}(z_{-})J'_{\ell}(z_{+})\Bigr]=0.          \label{54}
\end{equation}
As it follows from eq. (\ref{51}) that $q^{(-)2}-q^{(+)2}=4$, we need
$q^{(-)}\geq 1$. For a macroscopic vessel, $R\gg 1$, and $\ell<R$, the
cylindrical functions $I_{\ell}$ and $I'_{\ell}$ in the last equation can be
taken in asymptotics. This leads to
\begin{equation}
{\cal E}_{k\ell}q^{(-)}_{k\ell}(q^{(+)2}_{k\ell}+q^{(-)2}_{k\ell})J'_{\ell}(z_{+})
-\Gamma\,\frac{\ell}{R}\,(q^{(-)3}_{k\ell}J_{\ell}(z_{+})+q^{(+)2}_{k\ell})=0.
                                                       \label{55}
\end{equation}

In the irrotational case, $\Gamma=0$, or for $\ell=0$, the
dispersion equation (\ref{55}) determines the usual hydrodynamic
spectrum $z=z^{\circ}_{k\ell}$, where
$J'_{\ell}(z^{\circ}_{k\ell})=0$. Assuming a small shift $\delta
z_{k\ell}$ of eigenvalues, we find (omitting subscripts $k,\ell$)
\begin{equation}
\frac{\delta z}{z^{\circ}}\approx \frac{4R^{2}+(z^{\circ})^{2}}
{4R^{2}+2(z^{\circ})^{2}}\sqrt{\frac{\xi}{\xi+(z^{\circ})^{2}+(z^{\circ})^{4}/4R^{2}}}
\,\frac{\ell}{\ell^{2}+(z^{\circ})^{2}},                   \label{56}
\end{equation}
where the parameter $\xi$ was used earlier, eq. (\ref{25}). This
result is valid for short wavelengths, when $q^{(+)}\sim
q^{(-)}\sim 1$, $z^{\circ}\sim R\sim \ell$, and $\delta
z/z^{\circ}\sim \Gamma/\ell \ll 1$, but also in the long
wavelength limit, $q^{(+)}\ll 1$, if $\ell$ is sufficiently large,
$1\ll \ell\sim q^{(+)}R\ll R$. In the last case it is applicable
even for $\xi\sim(z^{\circ})^{2}$ since $\delta z/z^{\circ}\sim
\ell^{-1}\ll 1$. For excitations with small values of quantum
numbers $\ell,k \sim 1$ and $\Gamma\sim 1/R$, the displacements of
the roots is significant, and one has to return to the general eq.
(\ref{54}). In that case, $\Gamma\sim q^{(+)}$, and the presence
of the gap in the spectrum (\ref{50}) is essential.

The simplest way to finding the normalization constants is based on the
orthogonality conditions following from the set of equations (\ref{38}-\ref{40}),
\begin{equation}
\int_{0}^{R}r\,dr\left\{R^{(0)}_{k'\ell}R^{(0)}_{k\,\pm
\ell}\pm\Phi^{(0)}_{k'\ell} \Phi^{(0)}_{k,\,\pm \ell}\pm
F^{(0)}_{k'\ell}P^{(0)}_{k,\,\pm\ell}\right\} =\,\frac{1\pm
1}{2}\,\Theta_{k\ell}\delta_{kk'},                 \label{57}
\end{equation}
where the right hand side quantities $\Theta_{k\ell}$, after some algebra using
the equations and boundary conditions, can be expressed as
\begin{equation}
\Theta_{k\ell}=-\,\frac{\Gamma \ell}{D_{k\ell}{\cal E}_{k\ell}}\Bigl
[P^{(0)}_{k\ell}(R)\Bigr]^{2}+\,\frac{2{\cal E}_{k\ell}^{2}}{D_{k\ell}}\,
\int_{0}^{R}dr\,rF^{(0)}_{k\ell}P^{(0)}_{k\ell}.                 \label{58}
\end{equation}
On the other hand, conditions (\ref{57}) and operator expansions (\ref{37})
show that
\begin{equation}
A_{k\ell}=\,\frac{\sqrt{2}}{\Theta_{k\ell}}\int_{0}^{R}dr\,r\left\{R^{(0)}_{k\ell}
\hat{v}_{r}^{(\ell)}-i\Phi^{(0)}_{k\ell}\hat{v}^{(\ell)}_{\varphi}
-iP^{(0)}_{k\ell}\hat{n}^{(\ell)}\right\}.           \label{59}
\end{equation}

The quantization according to the Bose-statistics,
$[A_{k\ell},A^{\dagger}_{k'\ell'}]=\delta_{kk'}\delta_{\ell\ell'}$, together
with the original rules of quantum hydrodynamics (\ref{7}) transformed in our
problem to eq. (\ref{14}), leads in this approximation to
\begin{equation}
\frac{4}{\Theta^{2}_{k\ell}}\,\int_{0}^{R}dr\,r\left\{\Gamma R^{(0)}_{k\ell}
\Phi^{(0)}_{k\ell}+{\cal E}_{k\ell}\Bigl[(R^{(0)}_{k\ell})^{2}+
(\Phi^{(0)}_{k\ell})^{2}\Bigr]\right\}=1.              \label{60}
\end{equation}
The explicit calculation of the integral (\ref{60}) and comparison to eq.
(\ref{58}) determines the simple normalization condition,
\begin{equation}
\Theta_{k\ell}=2{\cal E}_{k\ell}.                      \label{61}
\end{equation}
The same result can be derived by the direct diagonalization of the linearized
Hamiltonian. We will not write down a cumbersome expression for the
normalization constants $C_{k\ell}$ that can be derived from eqs. (\ref{58})
and (\ref{61}) with the aid of the solutions (\ref{49}) and (\ref{52}). Here we
show only the result for a pure hydrodynamic spectrum that becomes exact for
$\ell=0$, when $b^{(-)}_{k\ell}\rightarrow 0$, while
\begin{equation}
(b^{(+)}_{k\ell})^{2}\approx
8D_{k\ell}\left\{(q^{(-)}_{k\ell})^{2}J_{\ell}^{2}(q^{(+)}_{k\ell}R)\left[
{\cal E}_{k\ell}R^{2}\left(1-\frac{\ell^{2}}{(q^{(+)}_{k\ell}R)^{2}}\right)
-\frac{\Gamma\ell(q^{(+)}_{k\ell})^{2}} {4{\cal
E}_{k\ell}^{2}}\right]\right\}^{-1}.                       \label{62}
\end{equation}

In spite of the presence of the gap in the energy spectrum (\ref{50}), we still
obtain in the transition to an infinite system, $R\rightarrow\infty$, a
vanishing interaction strength with the long-wavelength, $q^{(+)}\rightarrow
0$, phonons. Indeed, in agreement with general arguments, we see from eq.
(\ref{62}) that
\begin{equation}
b^{(+)}_{k\ell}\sim \sqrt{D_{k\ell}}\propto q^{(+)}_{k\ell}. \label{63}
\end{equation}
It is easy to see also that the off-diagonal matrix elements of the local
density are small being at $\Gamma\sim q^{(+)}\sim 1/R$ of the order $N^{-1/2}$
as it should be for collective excitations,
\begin{equation}
\langle k\ell|\hat{n}(r)|0\rangle\sim \sqrt{\frac{q^{(+)}_{k\ell}}{N}}\,
\frac{J_{\ell}(q^{(+)}_{k\ell}r)}{J_{\ell}(q^{(+)}_{k\ell}R)}.  \label{64}
\end{equation}
This justifies the linearization similar to the random phase approximation.

Our energy spectrum (\ref{50}) formally allows for two types of excitations
with phonon angular momentum $\ell$ and energies
\begin{equation}
E^{\pm}_{k\ell}=\,\frac{\ell\Gamma}{2}\,\pm\sqrt{\Gamma^{2}+(q^{(+)}_{k\ell})^{2}
+(q^{(+)}_{k\ell})^{4}/4}.                                \label{65}
\end{equation}
In the long wavelength limit, the Coriolis interaction, $-\ell\Omega$, of the
``minus" phonon with $\ell>2$ is so strong that the creation of such a phonon
becomes energetically favorable at sufficiently large angular velocity, namely
if $\Gamma>2\{[(q^{(+)}_{k\ell})^{2}+(q^{(+)}_{k\ell})^{4}/4]/(\ell^{2}-4)\}$.
However, this would correspond to the supersonic velocity of rotation at the
boundary, $\Gamma R/2>1$, when the approximation used above is invalid and the
spatial non-uniformity is essential. In our approximation we have to take into
account only the ``plus" branch $E^{+}_{k\ell}$.

\section{Discussion}

The study of rotating helium II has a long history. Contrary to
the predictions of Landau theory\cite{landau41}, it appeared that
below the $\lambda$-point the superfluid components behaved in
rotation as a classical fluid. This was shown by measuring the
shape of the free surface meniscus of rotating helium that
revealed a classical profile
\cite{Osborne50,Andronikashvili55,Donnelly56,Donnelly58,Turkington63,Meservey64}.
The measurement of the angular momentum of rotating helium II was
also found to correspond to the rigid body rotation of the
liquid\cite{Hall55,Vinen57,Reppy60}. As is well known, this
paradox was solved by the Onsager-Feynman theory of quantized
vortices\cite{feynman55}. The critical velocity for the vortex
nucleation is $\Omega_s=(\hbar/mR^2)\ln(R/a)$, where $R$ is the
radius of the vessel and $a$ is the radius of the vortex core. In
the aforementioned experiments the speed of rotation was much
higher than $\Omega_s$ so that the vortex lines formed a lattice
that on average imitated the rigid-body rotation \cite{feynman55}.

On the other hand, we cannot exclude the manifestation of
``rigid-body" rotation in several reported experiments. For
example, the interpretation of some data in \cite{kojima71} as
Landau state does not seem unambiguous. In this paper the helium
was contained in a torus resonator filled with slip rings and
brushes so that the normal-fluid component in rotation was locked
to the resonator by its viscous interaction. The relative velocity
of the normal fluid and superfluid was determined by measuring the
frequency difference of fourth-sound modes which run around the
resonator in both directions. The dependence of the superfluid angular
velocity on the normal-fluid angular velocity had the form of a
hysteretic curve. The reversible part of this curve was attributed
to the vortex-free potential flow (Landau state). Plausibly,
this reversible part could be as well attributed to the ``rigid"
velocity field, since for the multiple-connected geometry the
contribution from such a rotation of superfluid (if it does
exist) cannot be distinguished from the contribution of
irrotational potential motion.

It is also appropriate to mention here the paper
\cite{walmsley58}, where the angular momentum of superfluid helium
was measured after the rotating helium bucket has been stopped.
The resulting angular momentum was smaller than the angular momentum
of rotating bucket, and the authors came to a conclusion
that the reason for this was the formation of the long-lived
macroscopic rotational state that accepted part of the angular
momentum. As the authors pointed out, this fact could not be
explained neither by the Landau two-fluid theory (in a simply
connected region, the only solution for potential flow of
incompressible fluid is $v=0$) nor by Onsager-Feynman vortex
states. In this connection it is worthwhile to remember the so
called Lin's hydrodynamics\cite{lin87}, where, to our knowledge,
for the first time the conception of rotational superfluid flow
has been proposed. Lin has suggested that the superfluid might
have very small nonzero viscosity and would thus reach a steady
state of uniform rotation in which boundary slip would cause the
superfluid to rotate more slowly than the normal fluid at low
velocities. Lin has applied his approach for the explanation of
the variety of helium-II experiments. In particular, the above
mentioned experiment \cite{walmsley58} has found a satisfactory
explanation within his theory\cite{lin59}.

For our velocity field ((\ref{19}) with $K=0$), it is easy to
calculate critical velocities for slow rotation assuming that the
density is constant over the volume. For the rotation in a
circular cylinder of radius $R$ we obtain, by minimizing the free
energy in a rotating frame, the critical angular velocities for
vortex-free rotation $\Omega_b=\Omega_0 (2p+1)$, where $p=0, 1,
2...$. Here $\Omega_0=\hbar/mR^2$, and (see (\ref{22})) $s=0$ for
$\Omega<\Omega_0$, $s=1$ for $\Omega_0<\Omega<3\Omega_0$, $s=2$
for $3\Omega_0<\Omega<5\Omega_0$ and so on. Comparing this with
the critical velocity for the appearance of the first vortex
$\Omega_s$, we see that the first critical velocity for
vortex-free rotation $\Omega_0$ is approximately ten times smaller
than the vortex critical velocity $\Omega_s$. In general, our
critical velocities are on the order of $\Omega_0$ which even for
$R\approx 1 $ mm are rather small, about $0.01$ rad/s.

Now that the instrumentation methods are greatly advanced as
compared to those used in early experiments, a possible
clear-cut experiment for a search of the frozen rotation would be
the measuring of the free surface of rotating helium with the
speed of rotation being much less than the critical velocity for
the vortex nucleation. Other possibility is the direct
investigation of the flow pattern below the $\lambda$-point with the
aid of some visualization technique (see\cite{Zhang05} and
references therein). It is also worthwhile to note the paper\cite{Ellis93}
where the circularly polarized third-sound wave modes were used to drive
persistent flow states in a superfluid helium film inside the
resonator. As was pointed in this paper, this method could, in
principle, be used to find the profile of the velocity field.

One can also point out a possible connection of our solution to
the effects emerging in the flow of helium-II in porous glass with
the size 1-10 nm of the pores \cite{mendelson47}, narrow
slits\cite{Bowers52}, in small tubes \cite{atkins51}, and
submicron channels \cite{zimm90, Varoquaux94}. In particular, as
was reported in \cite{zimm90}, the critical velocities for the
flow of superfluid $^4$He through a submicron aperture appeared to
be much smaller than it would be necessary for a vortex ring
nucleated at the wall.

An interesting question of experimental manifestations of a
``rigid-body" rotation of quantum Bose-fluids other than liquid
helium requires special consideration and is beyond the scope of
the present paper.

\section{Conclusion}

In this article we made an attempt to vindicate the old operator formulation
of quantum hydrodynamics due to Landau \cite{landau41} and Geilikman
\cite{geilikman54}. This formulation does not lead to any unphysical states
of the type of the "fall onto center" and allows one, in the simplest
cylindrical geometry, to write down and solve the operator equations of motion for a
quantum liquid. An example of the explicit solution was found in the form of
the classical background and collective oscillatory modes. The solution
in general contains a superposition of a vortex line of intensity $K$ that
should be quantized through its relation to the angular momentum and a rigid-body
rotation with a local angular velocity $\Omega(r)=\Gamma\nu(r)/2$ due to
the velocity curl ``frozen" into the density of the liquid.

We considered in detail the solutions without a vortex
singularity, $K=0$. The calculations are rather long but
straightforward. On top of such a solution we superimpose small
oscillations of transverse components of the velocity and related
oscillations of density and linearize the equations of motion. For
the spatial function we get a complete set of modes with orbital
momentum $\ell$ and the radial dependence determined by the
combination of cylindrical functions $J_{\ell}(kr)$ and
$I_{\ell}(kr)$ that is determined by the boundary conditions.

Now, when the new types of quantum liquids are under experimental
scrutiny, the search for diverse manifestations of rotational superfluid
flow is at the frontiers of current interest. In any case, it seems that
the conventional consideration of the vortices as the only long lived
objects in rotating quantum liquids is too restrictive.\\
\\

We acknowledge useful discussions with S.T. Belyaev. We also thank
the referee who pointed out the similarity between our frozen
rotation and the vorticity of Laughlin's state in fractional
quantum Hall effect.

The research was partly supported by the NSF grant PHY-0758099.

\newpage

\underline{{\bf Appendix. Minimization of the density functional}.}\\
\\

To find the ground state based on the rotational velocity profile,
\begin{equation}
u(r)=\,\frac{\Gamma}{2\pi r}\,N(r),                       \label{66}
\end{equation}
we have to minimize the functional depending on $\nu(r)$ and its
gradient $d\nu/dr$. The functional contains kinetic energy ${\cal
K}$, potential (interaction) energy ${\cal W}$, and quantum
potential ${\cal Q}$. Let us consider these terms one by one.

The kinetic term is
\begin{equation}
{\cal K}=\,\frac{m}{2}\,\int d^{2}r\,\nu u^{2}.      \label{67}
\end{equation}
Here we minimize directly taking into account that we restrict ourselves
by the class of states, where eq. (\ref{66}) is valid,
\begin{equation}
\delta{\cal K}=\,\frac{m\Gamma^{2}}{8\pi^{2}}\int \frac{d^{2}r}{r^{2}}\,
\left\{\delta \nu(r)N^{2}(r)+2\nu(r)N(r)\delta N(r)\right\}. \label{68}
\end{equation}
Here, using the step function $\Theta(x)$,
\begin{equation}
\delta N(r)=\int d^{2}r'\,\theta(r-r')\delta\nu(r'),    \label{69}
\end{equation}
and changing notation in the double integral $r\leftrightarrow r'$,
\begin{equation}
\delta{\cal K}=\frac{m\Gamma^{2}}{8\pi^{2}}\int d^{2}r\left\{\frac{N^{2}(r)}{r^{2}}\,
+2\int \frac{d^{2}r'}{r'^{2}}\,\theta(r'-r)\nu(r')N(r')\right\}\delta\nu(r).
                                                        \label{70}
\end{equation}

The potential term,
\begin{equation}
{\cal W}=\,\frac{1}{2}\,m^{2}g\int d^{2}r\,(\nu-n_{0})^{2}, \label{71}
\end{equation}
is trivial,
\begin{equation}
\delta{\cal W}=m^{2}g\int d^{2}r\,(\nu-n_{0})\delta\nu(r).     \label{72}
\end{equation}

The quantum term,
\begin{equation}
{\cal Q}=\,\frac{\hbar^{2}}{8m}\,\int d^{2}r\,\frac{(\nabla\nu)^{2}}{\nu},
                                                                \label{73}
\end{equation}
depends both on $\nu$ and $\nabla\nu$. The variation goes as in the Lagrangian
mechanics,
\begin{equation}
\delta{\cal Q}=\,\frac{\hbar^{2}}{8m}\,\int d^{2}r\,\delta{\cal L}(\nu,\nabla\nu)=
\,\frac{\hbar^{2}}{8m}\,\int d^{2}r\,\left\{\frac{\partial{\cal L}}{\partial \nu}\,\delta\nu+
\,\frac{\partial{\cal L}}{\partial(\nabla\nu)}\cdot\delta(\nabla\nu)\right\}. \label{74}
\end{equation}
Here
\begin{equation}
\frac{\partial{\cal L}}{\partial \nu}=-\,\frac{(\nabla\nu)^{2}}{\nu^{2}}, \label{75}
\end{equation}
\begin{equation}
\frac{\partial{\cal L}}{\partial(\nabla\nu)}=\frac{2}{\nu}\,\Bigl(\nabla\nu\cdot
\delta(\nabla \nu)\Bigr).                                     \label{76}
\end{equation}
Since the symbols $\delta$ and $\nabla$ commute, eq. (\ref{76}) gives
\begin{equation}
2\nabla\cdot\left(\frac{\nabla\nu}{\nu}\,\delta\nu\right)-2\left(\nabla\cdot
\frac{\nabla\nu}{\nu}\right)
=2\nabla\cdot\left(\frac{\nabla\nu}{\nu}\,\delta\nu\right)
-2\,\frac{\nabla^{2}\nu}{\nu}\,+2\,\frac{(\nabla\nu)^{2}}{\nu^{2}}. \label{77}
\end{equation}
Combining eqs. (\ref{75}) and (\ref{76}), we obtain
\begin{equation}
\delta{\cal Q}=\,\frac{\hbar^{2}}{8m}\int d^{2}r\,\left\{\frac{(\nabla\nu)^{2}}
{\nu^{2}}\,-2\,\frac{\nabla^{2}\nu}{\nu}\right\}\delta\nu+\delta{\cal Q}_{S},   \label{78}
\end{equation}
where we have also a surface term,
\begin{equation}
\delta{\cal Q}_{S}=\frac{\hbar^{2}}{4m}\,\int d^{2}r\,\nabla\left(\frac{\nabla\nu}{\nu}\,
\delta\nu\right).                                     \label{79}
\end{equation}

If we forget about the surface contribution, the Lagrange equation is
\begin{equation}
m^{2}g(\nu-n_{0})+\,\frac{\hbar^{2}}{8m}\,\left\{\frac{(\nabla\nu)^{2}}
{\nu^{2}}\,-2\,\frac{\nabla^{2}\nu}{\nu}\right\}        \label{80}
\end{equation}
\[=-\,\frac{m\Gamma^{2}}{8\pi^{2}}\left\{\frac{N^{2}(r)}{r^{2}}\,
+2\int \frac{d^{2}r'}{r'^{2}}\,\theta(r'-r)\nu(r')N(r')\right\}.\]
Taking the gradient $d/dr$ in eq. (\ref{80}) and using again eq. (\ref{66}),
we come to eq. (\ref{24}).

\end{document}